\documentclass[%
 reprint,
 amsmath,amssymb,
 aps,
]{revtex4-2}

\usepackage{graphicx}
\usepackage{dcolumn}
\usepackage{bm}
\usepackage{color}

\begin{document}

\preprint{APS/123-QED}
\title{Holographic entanglement entropy of two disjoint intervals in AdS$_3$/CFT$_2$}
\author{Jun Tsujimura}
\email{tsujimura.jun.m5@s.mail.nagoya-u.ac.jp}
\author{Yasusada Nambu}
\email{nambu@gravity.phys.nagoya-u.ac.jp}
\affiliation{Department of Physics, Nagoya University, Chikusa, Nagoya 464-8602, Japan}

\begin{abstract}
    The Ryu-Takayanagi conjecture predicts a holographic dual of the entanglement entropy of a CFT. It proposes that the entanglement entropy is given by the area of the minimal surface in the dual spacetime. In the semi-classical limit, this conjecture is supported by the saddle point approximation. If there are multiple classical solutions, it is assumed that only the minimal action contributes to the entanglement entropy. However, we will point out that these saddles equally contribute to the entanglement entropy in some cases. Therefore, the derivation of the conjecture is incomplete if there are multiple extremal surfaces that extend from a sub-system on the AdS boundary. We will consider two disjoint intervals in CFT$_{1+1}$ as the simplest but non-trivial example, and propose another candidate for a holographic dual of the entanglement entropy of this system, which is the sum of all the signed areas of extremal surfaces in the dual spacetime. After that, we will derive it from the CFT calculations and propose the corresponding gravity side action.
\end{abstract}
\maketitle

\section{\label{sec:level1}Introduction}

    The AdS/CFT correspondence \cite{Maldacena:1997re,Gubser:1998bc,Witten:1998qj} often translates complex problems of $d+1$-dimensional quantum field theory into simple problems of $d+2$-dimensional classical gravity theory. As an example, consider the problem to evaluate an entanglement entropy(EE). It is one of the measures of correlations in quantum theory. Since, in general, it is difficult to evaluate an EE in quantum field theory, the AdS/CFT correspondence may be a helpful computational tool for it. It is natural to consider constructing the holographic formula to calculate an EE of CFT. We call an EE calculated by a holographic formula as a holographic entanglement entropy(HEE). Recently, a holographic formula of an EE is regarded as a foundation for studying the holographic aspect of gravity theory \cite{Czech:2012bh,Swingle:2009bg,Pastawski:2015qua,Dong:2016eik}.

    The Ryu-Takayanagi conjecture \cite{Ryu:2006bv,Ryu:2006ef} proposes a candidate of an HEE of static holographic CFT in the semi-classical limit. The definition of the EE $S_A$ of a sub-system $A$ is as follows.
\begin{align}\label{eq:def_of_EE}
    S_A = -\mathrm{tr}_A \left[ \rho_A \log \rho_A \right],
\end{align}
    where $\mathrm{tr}_A$ and $\rho_A$ are the trace operation and the reduced density matrix on $A$, respectively. The Ryu-Takayanagi conjecture proposes that
\begin{align}
    S_A = \frac{\text{Area of }\gamma_A}{4G},
\end{align}
    where $G$ is the gravitational constant, and $\gamma_A$ is the globally minimal surfaces among the extremal surfaces that satisfy the homologous condition \cite{Headrick:2007km}. The global minimality and the homologous condition are imposed not to cause the contradiction that the HEE violates the strong sub-additivity inequality. The generalized gravitational entropy \cite{Lewkowycz:2013nqa} explains why extremal surfaces describe an EE by the saddle point approximation. This partially supports the statement of the Ryu-Takayanagi conjecture.

    In this paper, we will point out that the derivation of the Ryu-Takayanagi conjecture and the corresponding CFT calculation is incomplete if there are multiple extremal surfaces for a sub-system. As it is based on the replica trick and thus we perform a limit operation, we have to perform these operations carefully. To derive the Ryu-Takayanagi conjecture in the conventional method, we need to assume the deliberate gravity action. Notice that there exists room for modifying the Ryu-Takayanagi conjecture since the global minimality and the homologous condition are just sufficient conditions for the HEE not to contradict the strong sub-additivity. Thus, we will consider another candidate of the HEE of two disjoint intervals for $1+1$ dimensional CFT.

    This paper is organized as follows. In Section \ref{sec:2}, we will apply the Ryu-Takayanagi conjecture to two disjoint intervals in $1+1$-dimensional CFT. Then, we will review the derivation of it and point out that we have to assume the specific gravity action to derive it. In Section \ref{sec:3}, we will try to construct another candidate of the above HEE by dealing with all extremal surfaces in the bulk and show that only one combination of the extremal surfaces with the plus or minus sign does not have a contradiction against the positivity and the sub-additivity of the EE. In Section \ref{sec:4}, we derive the EE from the CFT calculation and we propose a corresponding gravity action that yields the above HEE. Section \ref{sec:5} is the conclusion of the paper.

\section{\label{sec:2} Ryu-Takayanagi conjecture for two disjoint intervals}

\begin{figure}[t]
    \includegraphics[width=0.9\linewidth]{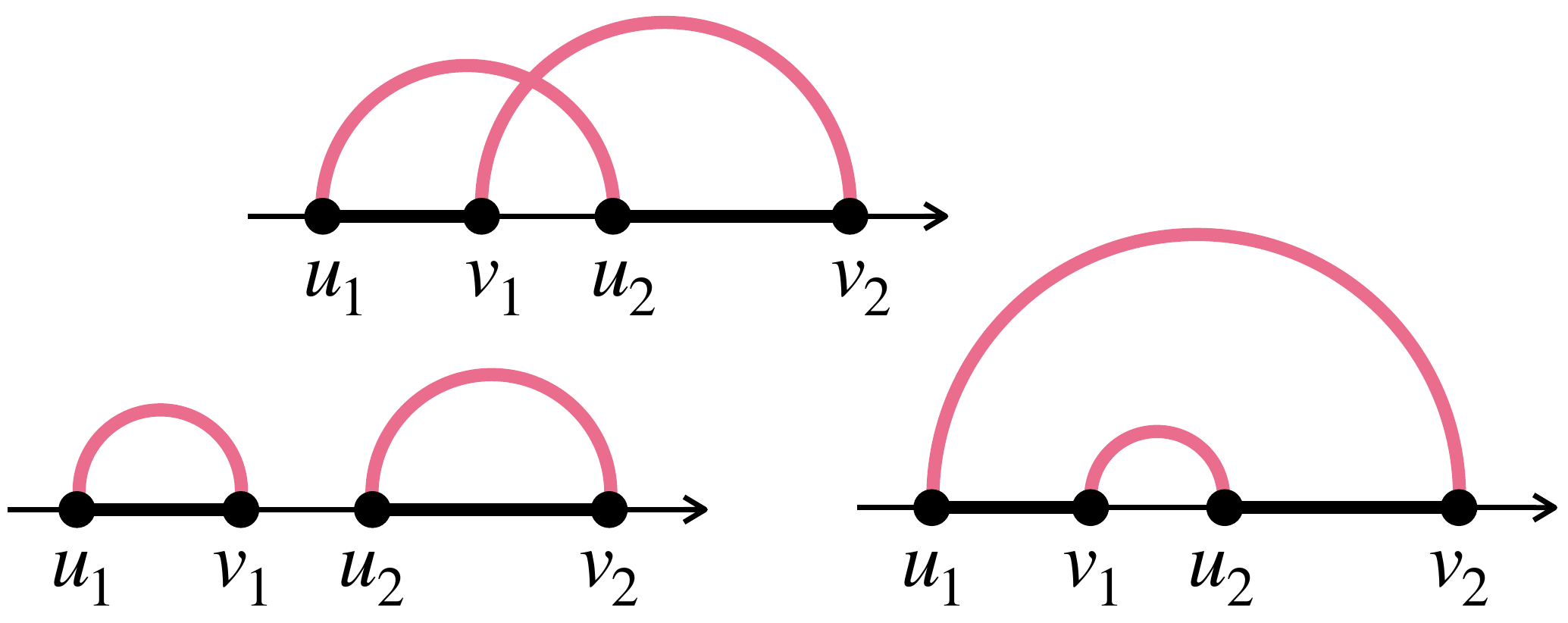}
    \caption{
        \label{fig:RTtwointervals} The magenta lines denote the three extremal surfaces $\gamma_A^{(i)},\ (i=1,2,3)$ satisfying the homologous condition of the two disjoint intervals $A$. The Ryu-Takayanagi conjecture proposes that the EE of $A$ is given by the length of the lines that has the minimal length in the above three combinations of the extremal surfaces.
    }
\end{figure}

    We will apply the Ryu-Takayanagi conjecture to two disjoint intervals and then review the derivation of it. Consider the large $c$ limit CFT on the $1+1$-dimensional Minkowski spacetime with a Cartesian coordinate $(x_0,x_1)$. Let $A$ denote the two disjoint intervals on a time slice $x_0=0$ defined as $A = A_1 + A_2$, $A_i= \left\{x_0, x_1\, |\, x_0=0, \, x_1 \in [u_i,v_i]\right\}$, where $v_2-u_2>\epsilon, u_2-v_1>\epsilon, v_1-u_1>\epsilon$ and $\epsilon$ is the UV cut-off scale of the CFT. Its dual spacetime is the AdS$_{2+1}$ spacetime described by the Poincar\'{e} patch. Figure~\ref{fig:RTtwointervals} shows a schematic picture of the above two disjoint intervals and the three combinations of the extremal surfaces $\gamma_A^{(i)},\ (i=1,2,3)$ satisfying the homologous condition that there exists the codimension $1$ region $\mathcal{R}^{(i)}$ such that $\partial \mathcal{R}^{(i)} = \gamma_A^{(i)} \cup A$. In this system, an extremal surface is a space-like geodesic and the EE of $A$ is given by their length. We will use terms for a surface and a line, and for an area and a length, interchangeably.
    Then, the Ryu-Takayanagi conjecture predicts the EE $S_A$ of the sub-system $A$ as follows:
\begin{align}
    S_A \label{eq:Ryu-Takayanagi conjecture}
    = \min_{i=1,2,3} & \left\{\frac{\text{Area of } \gamma_A^{(i)}}{4G} \right\} \\
    =\min &\left\{
    \frac{c}{3}\log\frac{v_1-u_1}{\epsilon}+\frac{c}{3}\log\frac{v_2-u_2}{\epsilon},\right.\nonumber\\ 
    &\ \ \, \frac{c}{3}\log\frac{u_2-u_1}{\epsilon}+\frac{c}{3}\log\frac{v_2-v_1}{\epsilon},\nonumber\\
    &\ \, \left. \frac{c}{3}\log\frac{v_2-u_1}{\epsilon}+\frac{c}{3}\log\frac{u_2-v_1}{\epsilon}
    \right\}
\end{align}
    where the gravitational constant $G$ is related to the central charge as $c = 3/2G$ from the Brown-Henneaux formula \cite{Brown:1986nw}.

    Let us see the derivation of the Ryu-Takayanagi conjecture. In the following, we implicitly consider the above two disjoint intervals system. From the replica trick, we can express the EE of a sub-system $A$ in terms of the partition functions as follows \cite{Calabrese:2009qy}:
\begin{align}
    S_A = \lim_{n \to 1} \frac{1}{1-n}\left( \log Z_A(n) - n \log Z \right),
\end{align}
    where $Z$ is the partition function and $Z_A(n)$ is the $n$-replicated partition function that the field operators are appropriately $n$-replicated around $\partial A$. If taking the semi-classical limit and there are $N$ classical solutions, the partition functions are described by the saddle point approximation as follows. 
\begin{gather}
    Z_A(n) = \sum_{i=1}^N p_i\, \exp\left[ - I_A^{(i)}(n) \right],\ 
    Z = \exp\left[- I \right],
\end{gather}
    where $I_A^{(i)}(n)$ and $I$ denote the classical action corresponding to the $i$-th saddle and the original one. The EE is expressed as 
\begin{gather}\label{eq:EE_derivation}
    S_A = \lim_{n \to 1}\frac{1}{1-n}\left( \log \sum_{i=1}^N p_i \exp\left[ -I_A^{(i)}(n) \right] + nI \right).
\end{gather}
    According to the AdS/CFT correspondence, we can use the gravity partition function to evaluate this instead of that of the CFT. Then, for example, $I$ is the Einstein-Hilbert action substituted by the metric tensor of the pure AdS spacetime if the dual CFT is the vacuum state. The actions $I_A^{(i)}(n)$ are assumed that $I_A^{(i)}(n) = n I +$  $(n-1)\gamma_A^{(i)}/4G + O((n-1)^2)$, where $\gamma_A^{(i)}$ is the areas of the $i$-th extremal surfaces that satisfy the homologous condition for $A$. For the above two disjoint intervals system, each $\gamma_A^{(i)}$ is the area of each of the three combinations of the extremal surfaces in the AdS spacetime as depicted in Figure~\ref{fig:RTtwointervals}. Then, we assume that only the saddle corresponding with the global minimal surfaces contributes to Eq.~\eqref{eq:EE_derivation}, and neglect the contributions of the other saddles. This is because the latter contributions to the path integral are exponentially smaller than the former's. Eventually, the EE may be described by only the global minimal surfaces with the homologous condition, and it seems that we can derive the Ryu-Takayanagi conjecture Eq.~\eqref{eq:Ryu-Takayanagi conjecture}.

    We should perform $n \to 1$ limit carefully for the above partition function if involving more than one saddle points. We can assume the effective action as $I_A^{(i)}(n) = n I + b^{(i)}_A(n)$, $\lim_{n \to 1}b^{(i)}_A(n) = 0$, then the first term of Eq.~\eqref{eq:EE_derivation} becomes 
\begin{align}\label{eq:effective action}
    \log &\left[ \sum_{i=1}^N p_i e^{-nI}\left( 1-b^{(i)}_A(n \to 1) \right) \right] \nonumber\\
    &= -nI + \log \sum_{i=1}^N p_i - \sum_{i=1}^N \frac{p_i b^{(i)}_A(n \to 1)}{ \sum_{i=1}^N p_i }
\end{align}
    Since the replica trick is based on $\lim_{n \to 1}\mathrm{tr} \rho^n_A = \lim_{n \to 1} Z_A(n)/Z^n = 1$, then $\sum_{i=1}^N p_i=1$ must hold. Thus, Eq.~\eqref{eq:EE_derivation} becomes
\begin{gather}
    S_A = \lim_{n \to 1} \sum_{i=1}^N \frac{p_i b^{(i)}_A(n)}{(n-1)}
\end{gather}
    Note that all saddles of the replicated gravity theory $Z_A(n)$ comparably contribute to the HEE, and then the above action does not derive the HEE expected by the Ryu-Takayanagi conjecture. Therefore, to derive the Ryu-Takayanagi conjecture in a system with multiple disjoint intervals, we should find another appropriate gravity action. For example, we need to assume the gravity action whose classical action contains only the global minimal surface, although it is almost tautology.

    We will not construct the gravity action from the CFT but focus on the hypothesis that the Ryu-Takayanagi conjecture is not valid for the system with multiple extremal surfaces. Remember that the global minimality and the homologous condition are not the unique way to choose which extremal surfaces contribute to an HEE, then there seems to be no positive reason why only a couple of the extremal surfaces describe an HEE. In the next section, we will provide another candidate of the HEE of the above two disjoint intervals, which is described by all extremal surfaces with the plus or minus sign.

\section{\label{sec:3} Another candidate of HEE for two disjoint intervals}

    We will neglect the global minimality and the homologous condition, and consider that the HEE of the above two disjoint intervals is described by all the six extremal surfaces depicted in Figure~\ref{fig:RTtwointervals}. Suppose that the HEE is the sum of the areas of all the extremal surfaces, then it has a contradiction because the resulting HEE violates the sub-additivity $S_{A} \le S_{A_1}+S_{A_2}$. Thus, we assume that some of the six surfaces contribute negatively to the HEE. We define the sign of the extremal surface in this sense. We will determine the sign of the extremal surfaces on the following assumptions. As the first assumption, the HEE is the sum of all signed areas of the extremal surfaces. The second assumption is the sign of each extremal surfaces depends on the topology of the sub-region and not on its size. With these two assumptions and from the consistency of the positivity of the EE and the mutual information, we will demonstrate that there only exists one combination of the signs of extremal surfaces describing the EE of the two disjoint intervals.

\begin{figure}[t]
    \includegraphics[width=0.7\linewidth]{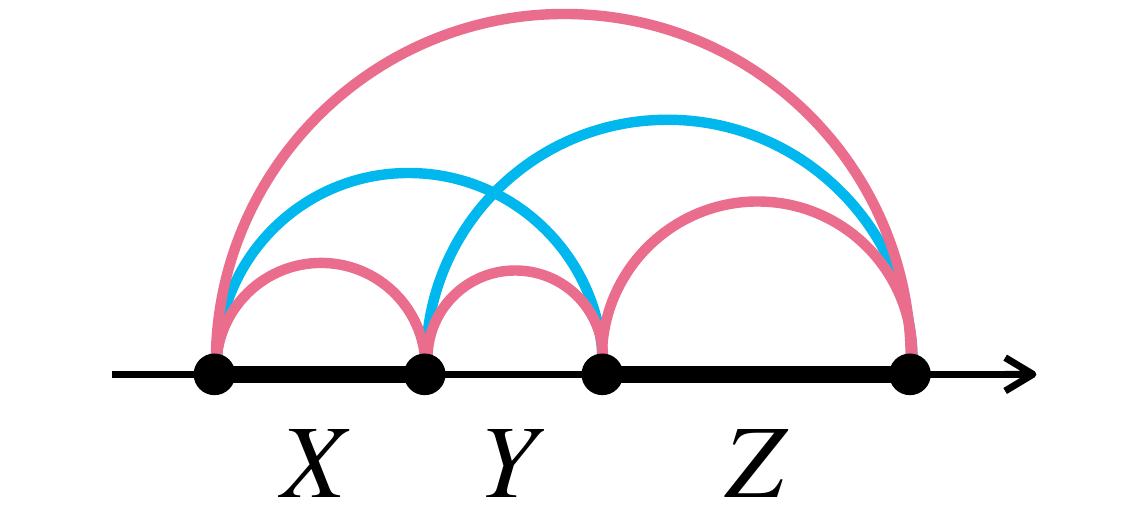}
    \caption{\label{fig:Thesurfaces} The unique acceptable combination of the extremal surfaces with the plus or minus sign. The magenta and cyan lines are the extremal surfaces with the plus and minus signs, respectively. }
\end{figure}

    Let $X$, $Y$ and $Z$ denote each three intervals $[u_1,v_1]$, $[v_1,u_2]$ and $[u_2,v_2]$ and each areas. There are the six local minimal surfaces that extend from the four points which are the boundary of the two disjoint intervals constituted from $X$ and $Z$. Let $(s_{X},\,s_{Y},\,s_{Z},\,s_{X+Y},\,s_{Y+Z},\,s_{X+Y+Z})$ denote the respective signs of extremal surfaces of the interval specified by its lower index. There exists the locally minimal surface uniquely that extends from two points at the AdS boundary to the AdS spacetime. Notice that the EE of the interval $X$ is $s_X = \frac{c}{3}\log{\frac{X}{\epsilon}}$ and then that of the other five in the same manner. In this notation, we will confirm that only $(+,+,+,-,-,+)$ is permitted as depicted in Figure~\ref{fig:Thesurfaces}.
    That is, the EE of the two disjoint intervals $X+Z$ should be
\begin{align}\label{eq:X+Z}
    S_{X+Z} = \frac{c}{3} \log \frac{(X+Y+Z)XYZ}{(X+Y)(Y+Z)\epsilon^2}.
\end{align}

    We will prove this statement with the above two assumptions and the positivity and the sub-additivity of the EE; $S_{X+Z}>0$ and $S_X + S_Z > S_{X+Z}$. First, let us consider the case that $X \gg Y,Z$. Two or three of $(s_{X},\,s_{X+Y},\,s_{X+Y+Z})$ should be the plus, otherwise $S_{X+Z} < 0$. On the other hand, if these three signs are the plus, the sub-additivity inequality $S_{X} + S_{Z} \ge S_{X+Z}$ is violated regardless of the signs of the others because $S_{X+Z} \sim 3 S_X$ becomes infinitely large compared with the others. Therefore, two of $(s_{X},\,s_{X+Y},\,s_{X+Y+Z})$ should be plus. As the same argument, two of $(s_{Z},\,s_{Y+Z},\,s_{X+Y+Z})$ should be plus.

    Next, consider the case that $Y \gg X,Z$. The number of the plus signs of $(s_{Y},\, s_{X+Y},\,s_{Y+Z},\,s_{X+Y+Z})$ is more than or equal to two, otherwise $S_{X+Z} < 0$. Conversely, if the number of the plus signs among them is three or four, the sub-additivity inequality $S_{X} + S_{Z} \ge S_{X+Z}$ is violated regardless of $(s_{X},\,s_{Z})$. To check this, consider the case $(-,+,-,+,+,-)$, which is the case that $S_{X+Z}$ takes the smallest value under the above condition. It violates the sub-additivity inequality because $\log \frac{Y(X+Y)(Y+Z)}{(X+Y+Z)XZ} > \log(XZ)$ under the condition $Y \gg X,Z$. Thus, $(s_{Y},\, s_{X+Y},\,s_{Y+Z},\,s_{X+Y+Z})$ needs to contains two each plus and minus signs.

    Table~\ref{tab:table1}.
\begin{table}[b]
\caption{\label{tab:table1}%
    The set of signs of extremal surfaces satisfying the positivity $S_{X+Z} \ge 0$ and the sub-additivity $S_{X} + S_{Z} \ge S_{X+Z}$ even if in the respective limit $X \gg Y,Z$, $Y \gg Z,X$ or $Z \gg X,Y$.
}
    \begin{ruledtabular}
        \begin{tabular}{cccccc}
        \textrm{$s_X$ }&
        \textrm{$s_Y$ }&
        \textrm{$s_Z$ }&
        \textrm{$s_{X+Y}$ }&
        \textrm{$s_{Y+Z}$ }&
        \textrm{$s_{X+Y+Z}$ }\\
        \colrule
        $+$ & $-$ & $+$ & $+$ & $+$ & $-$\\
        $-$ & $-$ & $+$ & $+$ & $-$ & $+$\\
        $+$ & $-$ & $-$ & $-$ & $+$ & $+$\\
        $+$ & $+$ & $+$ & $-$ & $-$ & $+$\\
        \end{tabular}
    \end{ruledtabular}
\end{table}
    shows the possible candidates for the set of signs from the discussion so far. Finally, we can confirm that these combinations except for $(+,+,+,-,-,+)$ have contradictions. The set of signs  $(+,-,+,+,+,-)$ in the first row of Table~\ref{tab:table1} contradicts to the sub-additivity $S_{X} + S_{Z} \ge S_{X+Z}$ due to the strong sub-additivity $S_{X+Y+Z}+S_{Y} \leq S_{X+Y} + S_{Y+Z}$. Note that its equality does not hold obviously. Next, consider the two sets of signs $(-,-,+,+,-,+)$ and $(+,-,-,-,+,+)$ in the second and the third rows of Table~\ref{tab:table1}. Let $X+Y+Z=\sqrt{2}\epsilon$, then the sub-additivity inequality yields to $\log 3 < \log 2$. Then, these combinations of the signs lead to the contradiction. Or taking the limit $X=Y \gg Z$, $(-,-,+,+,-,+)$ and $(+,-,-,-,+,+)$ lead $S_{X+Z}<0$ and $S_{X+Z}>S_X+S_Z$, respectively. Conversely, if $Y=Z \gg X$, the two sets of signs $(+,-,-,-,+,+)$ and $(-,-,+,+,-,+)$ lead $S_{X+Z}<0$ and $S_{X+Z}>S_X+S_Z$, respectively. Eventually, we conclude that only $(+,-,+,+,+,-)$ gives a possible candidate of the HEE.

    We restricted the combination of the signs of the extremal surfaces by only using the inequalities for the EE. In the same way, we can immediately predict the HEE of the $N$ disjoint intervals $A = \cup_{i=1}^N [u_i,v_i]$ as
\begin{align}
    S_A = \frac{c}{3}  \log \frac{\prod_{i, j}|v_i-u_j|}{\prod_{i<j}|u_i-u_j||v_i-v_j|} + \mathrm{const.}
\end{align}
    We can apply the above discussion to any system if its EE of one interval is the monotonically increasing function for its area. In addition to above these inequalities, we can use some other properties for an EE to restrict the signs of the surfaces. For example, $s_{X}, s_{Z}$ and $s_{X+Y+Z}$ should take plus sign by considering $Y,Z \to \epsilon$, $X,Y \to \epsilon$ or $Y \to \epsilon$ that are the one interval limit.

\section{\label{sec:4} Consistency and validity}

    We can derive Eq.~\eqref{eq:X+Z} from the CFT calculation \cite{Hartman:2013mia, Faulkner:2013yia}. Let the sub-system $A = [0,x]\cup[1,\infty],\ x=(v_1-u_1)(v_2-u_2)/(u_2-u_1)(v_2-v_1)$. We can evaluate the EE from the $4$-point correlation function.
\begin{align}
    S_A = \lim_{n \to 1} \frac{2}{1-n}\langle \mathcal{T}_n(0)\mathcal{\tilde{T}}_n(x)\mathcal{T}_n(1)\mathcal{\tilde{T}}_n(\infty) \rangle, 
\end{align}
    where $\mathcal{T}_n, \mathcal{\tilde{T}}_n$ are the holomorphic part of the twist operators with the conformal weight $h = c(n^2-1)/24n$. Let $c_2(x) = -\partial_x \log \langle \mathcal{T}_n(0)\mathcal{\tilde{T}}_n(x)\mathcal{T}_n(1)\mathcal{\tilde{T}}_n(\infty) \rangle$. For $n \sim 1$ in the semi-classical limit, we obtain two candidate of $c_2(x)$ as
\begin{align}
    c_2(x) = \frac{2h}{x-1},\ \frac{2h}{x}.
\end{align}
    What quantum state on the replica manifold is realized if there are two classical solutions? It should be expressed as $|\Omega\rangle = \sqrt{p_1} |\Omega_1\rangle  + \sqrt{p_2} |\Omega_2\rangle$, $\langle \Omega_1|\Omega_2\rangle \sim 0$, where $|\Omega_1\rangle$ and $|\Omega_2\rangle$ are the states corresponding with the respective classical solutions. Then, the partition function is described as $Z_A(n) = p_1 Z^{(1)}_A(n) + p_2 Z^{(1)}_A(n)$, where we defined $Z_A(n) = \langle \Omega|\Omega\rangle,\ Z^{(i)}_A(n) = \langle \Omega_i|\Omega_i\rangle$. Therefore, in Eq.~\eqref{eq:effective action}, we identify 
\begin{align}
    \frac{\partial}{\partial x} p_1 b^{(1)}_A(n) = \frac{2h}{x-1},\ \frac{\partial}{\partial x} p_2 b^{(2)}_A(n) = \frac{2h}{x}.
\end{align}
    From Eq.~\eqref{eq:effective action} and $\langle \mathcal{T}_n(u_1)\mathcal{\tilde{T}}_n(v_1)\mathcal{T}_n(u_2)\mathcal{\tilde{T}}_n(v_2) \rangle = (u_2-u_1)^{-2h}(v_2-v_1)^{-2h}\langle \mathcal{T}_n(0)\mathcal{\tilde{T}}_n(x)\mathcal{T}_n(1)\mathcal{\tilde{T}}_n(\infty) \rangle$, we can recover Eq.~\eqref{eq:X+Z}.

    How about replicated gravity action $I_A(n)$ that provides Eq.~\eqref{eq:X+Z}? As mentioned in \cite{Dong:2016fnf}, the replica manifold for CFT has conical singularities with angle $\Delta \phi = \pi(n-1/n)$ around $\partial A$. In the gravity side, $I_A(n)$ may contain cosmic strings with tension $\pm \Delta \phi/8\pi G$ since such cosmic strings create conical deficit angles with $\pm \Delta \phi$ around them, respectively. Then, consider the classical gravity action $I_A(n) = nI + I_{\text{strings}}(n) + I_A^{b.c.}(n)$, where $I_{\text{strings}}(n)$ is the cosmic strings action that describes the six cosmic strings that each string extends from $\partial A$, and $I_A^{b.c.}(n)$ imposes the boundary condition that the cosmic strings create conical deficit angle $\Delta \phi$ at each $\partial A$. In the above two disjoint intervals system, $I_{\text{strings}}(n)$ should contain four positive signed strings and two negative signed strings as depicted in Figure~\ref{fig:Thesurfaces}. If this classical action is the unique classical solution, then we can derive Eq.~\eqref{eq:X+Z} from this action substituting it into Eq.~\eqref{eq:EE_derivation}. Since the classical solution of the dual CFT is obtained on the same replica manifold, it is consistent that the bulk geometry is uniquely determined. As Eq.~\eqref{eq:X+Z} gives a smaller HEE than that from the Ryu-Takayanagi conjecture, this is the dominant saddle even if the conventional derivation of the Ryu-Takayanagi conjecture is valid. In the general system, the replicated gravity action $I_A(n)$ should be given by a bunch of the signed cosmic strings as well. Then, we have to clarify the condition to determine each sign of them instead of the homologous condition, although we do not consider about it in the paper. We may be able to derive it from the topological consistency of the replica manifolds between the gravity side and the CFT side \cite{Haehl:2014zoa}.

    We will focus on two outstanding properties of the HEE given by Eq.~\eqref{eq:X+Z} compared with the Ryu-Takayanagi conjecture. First, the mutual information $I(X:Z) = S_X+S_Z-S_{X+Z}$ from Eq.~\eqref{eq:X+Z} does not have a phase transition-like behavior. In the $Y \gg X,Z$ limit, Eq.~\eqref{eq:X+Z} leads to the following mutual information.
\begin{align}
    I(X:Z) = \frac{c}{3}\log \frac{(X+Y)(Y+Z)}{(X+Y+Z)Y} \longrightarrow \frac{c}{3} \frac{X Z}{Y^2}
\end{align}
    This is a $O(c^1)$ quantity, although the mutual information from the Ryu-Takayanagi conjecture is $O(c^0)$. The Ryu-Takayanagi conjecture predicts a qualitative change of the correlation in the system since the above mutual information measures the correlation between the sub-systems $X$ and $Z$. Since the correlation functions of CFT do not have a phase transition-like behavior as we can understand it from the GKP-Witten relation, it seems natural that an HEE also does not have such behavior at least in the large $c$ limit. Second, Eq.~\eqref{eq:X+Z} accidentally corresponds to the EE of the Luttinger liquid CFT with an appropriate parameter \cite{Calabrese:2009ez}. That is, Eq.~\eqref{eq:X+Z} is well behaved as an EE of a CFT, and then it deserves to study whether Eq.~\eqref{eq:X+Z} is the correct HEE of the two disjoint intervals.

\section{\label{sec:5} Conclusion}

    In summary, we should consider all the classical solutions for EE if there are multiple classical solutions, and we must assume an deliberate gravity action to derive the Ryu-Takayanagi conjecture in general. Instead, we showed another candidate of the HEE of the two disjoint intervals on some assumption instead of the global minimality and the homologous condition. The assumptions are that all the extremal surfaces in the bulk contribute to an HEE and that some of them contribute negatively to it. Then, we determined what combination of the signs of extremal surfaces acceptable is from the inequalities about the EE, and we derived it from the CFT calculation. Notice that the argument in Sec.\ref{sec:3} is robust in the sense that we can use the same logic to another system as long as the EE of one interval is strictly monotonically increasing function with respect to the area of it. After that, we showed a corresponding gravity action which is the action of the cosmic strings satisfying the deficit angle consistency.

    How to understand the entanglement wedge and the entangling surface if Eq.~\eqref{eq:X+Z} is the true HEE? As pointed out in \cite{Terashima:2020uqu}, we may need to reinterpret the conventional entanglement wedge, then Eq.~\eqref{eq:X+Z} may provide an alternative entanglement wedge with preferable properties. It may be possible to consider that the negative extremal surfaces make up for the correlation too much cut by the other positive extremal surfaces.

    In future work, we have to derive the action which provides an HEE from only the properties of the replicated CFT regardless of whether the Ryu-Takayanagi conjecture predicts the correct HEE or not. If Eq.~\eqref{eq:X+Z} is the correct EE, we have to reconsider arguments based on the global minimality and the homologous condition of the Ryu-Takayanagi conjecture.

\begin{acknowledgments}
    I would like to thank Seiji Terashima and Takato Mori for useful discussions and comments. This work was financially supported by JST SPRING, Grant Number JPMJSP2125 and by JSPS Fellows (22J14390). The author J.T. would like to take this opportunity to thank the "Interdisciplinary Frontier Next-Generation Researcher Program of the Tokai Higher Education and Research System. The author Y.N. was supported in part by JSPS KAKENHI Grant No. 19K03866.
\end{acknowledgments}

\bibliography{HEEforTwointervals}

\begin{thebibliography}{19}%
\makeatletter
\providecommand \@ifxundefined [1]{%
 \@ifx{#1\undefined}
}%
\providecommand \@ifnum [1]{%
 \ifnum #1\expandafter \@firstoftwo
 \else \expandafter \@secondoftwo
 \fi
}%
\providecommand \@ifx [1]{%
 \ifx #1\expandafter \@firstoftwo
 \else \expandafter \@secondoftwo
 \fi
}%
\providecommand \natexlab [1]{#1}%
\providecommand \enquote  [1]{``#1''}%
\providecommand \bibnamefont  [1]{#1}%
\providecommand \bibfnamefont [1]{#1}%
\providecommand \citenamefont [1]{#1}%
\providecommand \href@noop [0]{\@secondoftwo}%
\providecommand \href [0]{\begingroup \@sanitize@url \@href}%
\providecommand \@href[1]{\@@startlink{#1}\@@href}%
\providecommand \@@href[1]{\endgroup#1\@@endlink}%
\providecommand \@sanitize@url [0]{\catcode `\\12\catcode `\$12\catcode
  `\&12\catcode `\#12\catcode `\^12\catcode `\_12\catcode `\%12\relax}%
\providecommand \@@startlink[1]{}%
\providecommand \@@endlink[0]{}%
\providecommand \url  [0]{\begingroup\@sanitize@url \@url }%
\providecommand \@url [1]{\endgroup\@href {#1}{\urlprefix }}%
\providecommand \urlprefix  [0]{URL }%
\providecommand \Eprint [0]{\href }%
\providecommand \doibase [0]{https://doi.org/}%
\providecommand \selectlanguage [0]{\@gobble}%
\providecommand \bibinfo  [0]{\@secondoftwo}%
\providecommand \bibfield  [0]{\@secondoftwo}%
\providecommand \translation [1]{[#1]}%
\providecommand \BibitemOpen [0]{}%
\providecommand \bibitemStop [0]{}%
\providecommand \bibitemNoStop [0]{.\EOS\space}%
\providecommand \EOS [0]{\spacefactor3000\relax}%
\providecommand \BibitemShut  [1]{\csname bibitem#1\endcsname}%
\let\auto@bib@innerbib\@empty
\bibitem [{\citenamefont {Maldacena}(1999)}]{Maldacena:1997re}%
  \BibitemOpen
  \bibfield  {author} {\bibinfo {author} {\bibfnamefont {J.~M.}\ \bibnamefont
  {Maldacena}},\ }\bibfield  {title} {\bibinfo {title} {{The Large N limit of
  superconformal field theories and supergravity}},\ }\href
  {https://doi.org/10.1023/A:1026654312961} {\bibfield  {journal} {\bibinfo
  {journal} {Int. J. Theor. Phys.}\ }\textbf {\bibinfo {volume} {38}},\
  \bibinfo {pages} {1113} (\bibinfo {year} {1999})},\ \Eprint
  {https://arxiv.org/abs/hep-th/9711200} {arXiv:hep-th/9711200} \BibitemShut
  {NoStop}%
\bibitem [{\citenamefont {Gubser}\ \emph {et~al.}(1998)\citenamefont {Gubser},
  \citenamefont {Klebanov},\ and\ \citenamefont {Polyakov}}]{Gubser:1998bc}%
  \BibitemOpen
  \bibfield  {author} {\bibinfo {author} {\bibfnamefont {S.}~\bibnamefont
  {Gubser}}, \bibinfo {author} {\bibfnamefont {I.~R.}\ \bibnamefont
  {Klebanov}},\ and\ \bibinfo {author} {\bibfnamefont {A.~M.}\ \bibnamefont
  {Polyakov}},\ }\bibfield  {title} {\bibinfo {title} {{Gauge theory
  correlators from noncritical string theory}},\ }\href
  {https://doi.org/10.1016/S0370-2693(98)00377-3} {\bibfield  {journal}
  {\bibinfo  {journal} {Phys. Lett. B}\ }\textbf {\bibinfo {volume} {428}},\
  \bibinfo {pages} {105} (\bibinfo {year} {1998})},\ \Eprint
  {https://arxiv.org/abs/hep-th/9802109} {arXiv:hep-th/9802109} \BibitemShut
  {NoStop}%
\bibitem [{\citenamefont {Witten}(1998)}]{Witten:1998qj}%
  \BibitemOpen
  \bibfield  {author} {\bibinfo {author} {\bibfnamefont {E.}~\bibnamefont
  {Witten}},\ }\bibfield  {title} {\bibinfo {title} {{Anti-de Sitter space and
  holography}},\ }\href {https://doi.org/10.4310/ATMP.1998.v2.n2.a2} {\bibfield
   {journal} {\bibinfo  {journal} {Adv. Theor. Math. Phys.}\ }\textbf {\bibinfo
  {volume} {2}},\ \bibinfo {pages} {253} (\bibinfo {year} {1998})},\ \Eprint
  {https://arxiv.org/abs/hep-th/9802150} {arXiv:hep-th/9802150} \BibitemShut
  {NoStop}%
\bibitem [{\citenamefont {Czech}\ \emph {et~al.}(2012)\citenamefont {Czech},
  \citenamefont {Karczmarek}, \citenamefont {Nogueira},\ and\ \citenamefont
  {Van~Raamsdonk}}]{Czech:2012bh}%
  \BibitemOpen
  \bibfield  {author} {\bibinfo {author} {\bibfnamefont {B.}~\bibnamefont
  {Czech}}, \bibinfo {author} {\bibfnamefont {J.~L.}\ \bibnamefont
  {Karczmarek}}, \bibinfo {author} {\bibfnamefont {F.}~\bibnamefont
  {Nogueira}},\ and\ \bibinfo {author} {\bibfnamefont {M.}~\bibnamefont
  {Van~Raamsdonk}},\ }\bibfield  {title} {\bibinfo {title} {{The Gravity Dual
  of a Density Matrix}},\ }\href
  {https://doi.org/10.1088/0264-9381/29/15/155009} {\bibfield  {journal}
  {\bibinfo  {journal} {Class. Quant. Grav.}\ }\textbf {\bibinfo {volume}
  {29}},\ \bibinfo {pages} {155009} (\bibinfo {year} {2012})},\ \Eprint
  {https://arxiv.org/abs/1204.1330} {arXiv:1204.1330 [hep-th]} \BibitemShut
  {NoStop}%
\bibitem [{\citenamefont {Swingle}(2012)}]{Swingle:2009bg}%
  \BibitemOpen
  \bibfield  {author} {\bibinfo {author} {\bibfnamefont {B.}~\bibnamefont
  {Swingle}},\ }\bibfield  {title} {\bibinfo {title} {{Entanglement
  Renormalization and Holography}},\ }\href
  {https://doi.org/10.1103/PhysRevD.86.065007} {\bibfield  {journal} {\bibinfo
  {journal} {Phys. Rev. D}\ }\textbf {\bibinfo {volume} {86}},\ \bibinfo
  {pages} {065007} (\bibinfo {year} {2012})},\ \Eprint
  {https://arxiv.org/abs/0905.1317} {arXiv:0905.1317 [cond-mat.str-el]}
  \BibitemShut {NoStop}%
\bibitem [{\citenamefont {Pastawski}\ \emph {et~al.}(2015)\citenamefont
  {Pastawski}, \citenamefont {Yoshida}, \citenamefont {Harlow},\ and\
  \citenamefont {Preskill}}]{Pastawski:2015qua}%
  \BibitemOpen
  \bibfield  {author} {\bibinfo {author} {\bibfnamefont {F.}~\bibnamefont
  {Pastawski}}, \bibinfo {author} {\bibfnamefont {B.}~\bibnamefont {Yoshida}},
  \bibinfo {author} {\bibfnamefont {D.}~\bibnamefont {Harlow}},\ and\ \bibinfo
  {author} {\bibfnamefont {J.}~\bibnamefont {Preskill}},\ }\bibfield  {title}
  {\bibinfo {title} {{Holographic quantum error-correcting codes: Toy models
  for the bulk/boundary correspondence}},\ }\href
  {https://doi.org/10.1007/JHEP06(2015)149} {\bibfield  {journal} {\bibinfo
  {journal} {JHEP}\ }\textbf {\bibinfo {volume} {06}},\ \bibinfo {pages}
  {149}},\ \Eprint {https://arxiv.org/abs/1503.06237} {arXiv:1503.06237
  [hep-th]} \BibitemShut {NoStop}%
\bibitem [{\citenamefont {Dong}\ \emph {et~al.}(2016)\citenamefont {Dong},
  \citenamefont {Harlow},\ and\ \citenamefont {Wall}}]{Dong:2016eik}%
  \BibitemOpen
  \bibfield  {author} {\bibinfo {author} {\bibfnamefont {X.}~\bibnamefont
  {Dong}}, \bibinfo {author} {\bibfnamefont {D.}~\bibnamefont {Harlow}},\ and\
  \bibinfo {author} {\bibfnamefont {A.~C.}\ \bibnamefont {Wall}},\ }\bibfield
  {title} {\bibinfo {title} {{Reconstruction of Bulk Operators within the
  Entanglement Wedge in Gauge-Gravity Duality}},\ }\href
  {https://doi.org/10.1103/PhysRevLett.117.021601} {\bibfield  {journal}
  {\bibinfo  {journal} {Phys. Rev. Lett.}\ }\textbf {\bibinfo {volume} {117}},\
  \bibinfo {pages} {021601} (\bibinfo {year} {2016})},\ \Eprint
  {https://arxiv.org/abs/1601.05416} {arXiv:1601.05416 [hep-th]} \BibitemShut
  {NoStop}%
\bibitem [{\citenamefont {Ryu}\ and\ \citenamefont
  {Takayanagi}(2006{\natexlab{a}})}]{Ryu:2006bv}%
  \BibitemOpen
  \bibfield  {author} {\bibinfo {author} {\bibfnamefont {S.}~\bibnamefont
  {Ryu}}\ and\ \bibinfo {author} {\bibfnamefont {T.}~\bibnamefont
  {Takayanagi}},\ }\bibfield  {title} {\bibinfo {title} {{Holographic
  derivation of entanglement entropy from AdS/CFT}},\ }\href
  {https://doi.org/10.1103/PhysRevLett.96.181602} {\bibfield  {journal}
  {\bibinfo  {journal} {Phys. Rev. Lett.}\ }\textbf {\bibinfo {volume} {96}},\
  \bibinfo {pages} {181602} (\bibinfo {year} {2006}{\natexlab{a}})},\ \Eprint
  {https://arxiv.org/abs/hep-th/0603001} {arXiv:hep-th/0603001} \BibitemShut
  {NoStop}%
\bibitem [{\citenamefont {Ryu}\ and\ \citenamefont
  {Takayanagi}(2006{\natexlab{b}})}]{Ryu:2006ef}%
  \BibitemOpen
  \bibfield  {author} {\bibinfo {author} {\bibfnamefont {S.}~\bibnamefont
  {Ryu}}\ and\ \bibinfo {author} {\bibfnamefont {T.}~\bibnamefont
  {Takayanagi}},\ }\bibfield  {title} {\bibinfo {title} {{Aspects of
  Holographic Entanglement Entropy}},\ }\href
  {https://doi.org/10.1088/1126-6708/2006/08/045} {\bibfield  {journal}
  {\bibinfo  {journal} {JHEP}\ }\textbf {\bibinfo {volume} {08}},\ \bibinfo
  {pages} {045}},\ \Eprint {https://arxiv.org/abs/hep-th/0605073}
  {arXiv:hep-th/0605073} \BibitemShut {NoStop}%
\bibitem [{\citenamefont {Headrick}\ and\ \citenamefont
  {Takayanagi}(2007)}]{Headrick:2007km}%
  \BibitemOpen
  \bibfield  {author} {\bibinfo {author} {\bibfnamefont {M.}~\bibnamefont
  {Headrick}}\ and\ \bibinfo {author} {\bibfnamefont {T.}~\bibnamefont
  {Takayanagi}},\ }\bibfield  {title} {\bibinfo {title} {{A Holographic proof
  of the strong subadditivity of entanglement entropy}},\ }\href
  {https://doi.org/10.1103/PhysRevD.76.106013} {\bibfield  {journal} {\bibinfo
  {journal} {Phys. Rev. D}\ }\textbf {\bibinfo {volume} {76}},\ \bibinfo
  {pages} {106013} (\bibinfo {year} {2007})},\ \Eprint
  {https://arxiv.org/abs/arXiv:0704.3719} {arXiv:arXiv:0704.3719 [hep-th]}
  \BibitemShut {NoStop}%
\bibitem [{\citenamefont {Lewkowycz}\ and\ \citenamefont
  {Maldacena}(2013)}]{Lewkowycz:2013nqa}%
  \BibitemOpen
  \bibfield  {author} {\bibinfo {author} {\bibfnamefont {A.}~\bibnamefont
  {Lewkowycz}}\ and\ \bibinfo {author} {\bibfnamefont {J.}~\bibnamefont
  {Maldacena}},\ }\bibfield  {title} {\bibinfo {title} {{Generalized
  gravitational entropy}},\ }\href {https://doi.org/10.1007/JHEP08(2013)090}
  {\bibfield  {journal} {\bibinfo  {journal} {JHEP}\ }\textbf {\bibinfo
  {volume} {08}},\ \bibinfo {pages} {090}},\ \Eprint
  {https://arxiv.org/abs/arXiv:1304.4926} {arXiv:arXiv:1304.4926 [hep-th]}
  \BibitemShut {NoStop}%
\bibitem [{\citenamefont {Brown}\ and\ \citenamefont
  {Henneaux}(1986)}]{Brown:1986nw}%
  \BibitemOpen
  \bibfield  {author} {\bibinfo {author} {\bibfnamefont {J.}~\bibnamefont
  {Brown}}\ and\ \bibinfo {author} {\bibfnamefont {M.}~\bibnamefont
  {Henneaux}},\ }\bibfield  {title} {\bibinfo {title} {{Central Charges in the
  Canonical Realization of Asymptotic Symmetries: An Example from
  Three-Dimensional Gravity}},\ }\href {https://doi.org/10.1007/BF01211590}
  {\bibfield  {journal} {\bibinfo  {journal} {Commun. Math. Phys.}\ }\textbf
  {\bibinfo {volume} {104}},\ \bibinfo {pages} {207} (\bibinfo {year}
  {1986})}\BibitemShut {NoStop}%
\bibitem [{\citenamefont {Calabrese}\ and\ \citenamefont
  {Cardy}(2009)}]{Calabrese:2009qy}%
  \BibitemOpen
  \bibfield  {author} {\bibinfo {author} {\bibfnamefont {P.}~\bibnamefont
  {Calabrese}}\ and\ \bibinfo {author} {\bibfnamefont {J.}~\bibnamefont
  {Cardy}},\ }\bibfield  {title} {\bibinfo {title} {{Entanglement entropy and
  conformal field theory}},\ }\href
  {https://doi.org/10.1088/1751-8113/42/50/504005} {\bibfield  {journal}
  {\bibinfo  {journal} {J. Phys. A}\ }\textbf {\bibinfo {volume} {42}},\
  \bibinfo {pages} {504005} (\bibinfo {year} {2009})},\ \Eprint
  {https://arxiv.org/abs/arXiv:0905.4013} {arXiv:arXiv:0905.4013
  [cond-mat.stat-mech]} \BibitemShut {NoStop}%
\bibitem [{\citenamefont {Hartman}(2013)}]{Hartman:2013mia}%
  \BibitemOpen
  \bibfield  {author} {\bibinfo {author} {\bibfnamefont {T.}~\bibnamefont
  {Hartman}},\ }\bibfield  {title} {\bibinfo {title} {{Entanglement Entropy at
  Large Central Charge}},\ }\href@noop {} {\  (\bibinfo {year} {2013})},\
  \Eprint {https://arxiv.org/abs/1303.6955} {arXiv:1303.6955 [hep-th]}
  \BibitemShut {NoStop}%
\bibitem [{\citenamefont {Faulkner}(2013)}]{Faulkner:2013yia}%
  \BibitemOpen
  \bibfield  {author} {\bibinfo {author} {\bibfnamefont {T.}~\bibnamefont
  {Faulkner}},\ }\bibfield  {title} {\bibinfo {title} {{The Entanglement Renyi
  Entropies of Disjoint Intervals in AdS/CFT}},\ }\href@noop {} {\  (\bibinfo
  {year} {2013})},\ \Eprint {https://arxiv.org/abs/1303.7221} {arXiv:1303.7221
  [hep-th]} \BibitemShut {NoStop}%
\bibitem [{\citenamefont {Dong}(2016)}]{Dong:2016fnf}%
  \BibitemOpen
  \bibfield  {author} {\bibinfo {author} {\bibfnamefont {X.}~\bibnamefont
  {Dong}},\ }\bibfield  {title} {\bibinfo {title} {{The Gravity Dual of Renyi
  Entropy}},\ }\href {https://doi.org/10.1038/ncomms12472} {\bibfield
  {journal} {\bibinfo  {journal} {Nature Commun.}\ }\textbf {\bibinfo {volume}
  {7}},\ \bibinfo {pages} {12472} (\bibinfo {year} {2016})},\ \Eprint
  {https://arxiv.org/abs/arXiv:1601.06788} {arXiv:arXiv:1601.06788 [hep-th]}
  \BibitemShut {NoStop}%
\bibitem [{\citenamefont {Haehl}\ \emph {et~al.}(2015)\citenamefont {Haehl},
  \citenamefont {Hartman}, \citenamefont {Marolf}, \citenamefont {Maxfield},\
  and\ \citenamefont {Rangamani}}]{Haehl:2014zoa}%
  \BibitemOpen
  \bibfield  {author} {\bibinfo {author} {\bibfnamefont {F.~M.}\ \bibnamefont
  {Haehl}}, \bibinfo {author} {\bibfnamefont {T.}~\bibnamefont {Hartman}},
  \bibinfo {author} {\bibfnamefont {D.}~\bibnamefont {Marolf}}, \bibinfo
  {author} {\bibfnamefont {H.}~\bibnamefont {Maxfield}},\ and\ \bibinfo
  {author} {\bibfnamefont {M.}~\bibnamefont {Rangamani}},\ }\bibfield  {title}
  {\bibinfo {title} {{Topological aspects of generalized gravitational
  entropy}},\ }\href {https://doi.org/10.1007/JHEP05(2015)023} {\bibfield
  {journal} {\bibinfo  {journal} {JHEP}\ }\textbf {\bibinfo {volume} {05}},\
  \bibinfo {pages} {023}},\ \Eprint {https://arxiv.org/abs/1412.7561}
  {arXiv:1412.7561 [hep-th]} \BibitemShut {NoStop}%
\bibitem [{\citenamefont {Calabrese}\ \emph {et~al.}(2009)\citenamefont
  {Calabrese}, \citenamefont {Cardy},\ and\ \citenamefont
  {Tonni}}]{Calabrese:2009ez}%
  \BibitemOpen
  \bibfield  {author} {\bibinfo {author} {\bibfnamefont {P.}~\bibnamefont
  {Calabrese}}, \bibinfo {author} {\bibfnamefont {J.}~\bibnamefont {Cardy}},\
  and\ \bibinfo {author} {\bibfnamefont {E.}~\bibnamefont {Tonni}},\ }\bibfield
   {title} {\bibinfo {title} {{Entanglement entropy of two disjoint intervals
  in conformal field theory}},\ }\href
  {https://doi.org/10.1088/1742-5468/2009/11/P11001} {\bibfield  {journal}
  {\bibinfo  {journal} {J. Stat. Mech.}\ }\textbf {\bibinfo {volume} {0911}},\
  \bibinfo {pages} {P11001} (\bibinfo {year} {2009})},\ \Eprint
  {https://arxiv.org/abs/0905.2069} {arXiv:0905.2069 [hep-th]} \BibitemShut
  {NoStop}%
\bibitem [{\citenamefont {Terashima}(2021)}]{Terashima:2020uqu}%
  \BibitemOpen
  \bibfield  {author} {\bibinfo {author} {\bibfnamefont {S.}~\bibnamefont
  {Terashima}},\ }\bibfield  {title} {\bibinfo {title} {{Bulk locality in the
  AdS/CFT correspondence}},\ }\href
  {https://doi.org/10.1103/PhysRevD.104.086014} {\bibfield  {journal} {\bibinfo
   {journal} {Phys. Rev. D}\ }\textbf {\bibinfo {volume} {104}},\ \bibinfo
  {pages} {086014} (\bibinfo {year} {2021})},\ \Eprint
  {https://arxiv.org/abs/2005.05962} {arXiv:2005.05962 [hep-th]} \BibitemShut
  {NoStop}%
\end{thebibliography}%

\end{document}